\documentclass[twocolumn,showpacs,preprintnumbers,superscriptaddress,amsmath,amssymb]{revtex4-1}

\usepackage{epsfig}
\usepackage{graphicx}
\usepackage{dcolumn}
\usepackage{bm}
\usepackage{times}
\usepackage{xcolor}

\begin{document}


\title{Crossover phenomena of percolation transition in evolution networks with hybrid attachment}
\author{Xiaolong Chen}
\affiliation{Web Sciences Center, University of Electronic
Science and Technology of China, Chengdu 611731, China}
\affiliation{Big Data Research Center, University of Electronic
Science and Technology of China, Chengdu 611731, China}
\author{Chun Yang}
\affiliation{School of Mathematical Science, University of Electronic Science and Technology of China, Chengdu 611731, China}
\author{Linfeng Zhong}
\affiliation{Web Sciences Center, University of Electronic
Science and Technology of China, Chengdu 611731, China}
\affiliation{Big Data Research Center, University of Electronic
Science and Technology of China, Chengdu 611731, China}
\author{Ming Tang} \email{tangminghan007@gmail.com}
\affiliation{Web Sciences Center, University of Electronic
Science and Technology of China, Chengdu 611731, China}

\date{\today}

\begin{abstract}
A first-order percolation transition, called explosive percolation, was recently discovered in evolution networks with random edge selection under a certain restriction. However, the network percolation with more realistic evolution mechanisms such as preferential attachment has not yet been concerned. We propose a tunable network percolation model by introducing a hybrid mechanism of edge selection into the Bohman-Frieze-Wormald (BFW) model, in which a parameter adjusts the relative weights between random and preferential selections. A large number of simulations indicate that there exist crossover phenomena of percolation transition by adjusting the parameter in the evolution processes. When the strategy of selecting a candidate edge is dominated by random selection, a single discontinuous percolation transition occurs. When a candidate edge is selected more preferentially based on node`s degree, the size of the largest component undergoes multiple discontinuous jumps, which exhibits a peculiar difference from the network percolation of random selection with a certain restriction. Besides, the percolation transition becomes continuous when the candidate edge is selected completely preferentially.
\end{abstract}

\maketitle

\textbf{Percolation is a pervasive concept in graph theory and statistics physics, which has a broad application in network science, epidemiology, and so on. Now it's a useful theoretical tool to study the large-scale connectivity of complex networks. It was known as a continuous percolation process of random networks before the product rule was introduced into the network evolution process by Achlioptas in 2009. In the percolation model proposed by Achlioptas, the size of the largest component jumps abruptly at a transition point, which is called explosive percolation, and it is considered as a first-order transition. Stimulated by the work of Achlioptas, a great amount of attention has been given to the research of the explosive percolation in evolution networks with random edge selection under a certain restriction. However, the percolation transition in a more realistic evolution process of networks is still lack of studying. Based on the classical Bohman-Frieze-Wormald model, we propose a percolation model in evolution process of networks with hybrid attachment. In our model, a tunable parameter is taken into account to adjust the way of selecting a candidate edge. Through numeric simulations, we find that there are crossover phenomena of percolation transition in the evolution process, which exhibit a peculiar difference from the classical Bohman-Frieze-Wormald network percolation. Specifically, when the strategy of selecting a candidate edge is dominated by random selection, there is only a single discontinuous jump of the largest component size. When a candidate edge is selected more preferentially based on node`s degree, the largest component grows with multiple discontinuous jumps. Besides, nearly completely preferential selection makes the size of the largest component increase continuously. Our work reveals the effect of hybrid attachment mechanism on the network percolation transition, and offer some insights into understanding the network percolation in a more realistic evolution process.}

\section{Introduction} \label{sec:intro}
Percolation is used to describe the movement and filtering of fluids through porous materials in materials science
\cite{grimmett1999percolation}. During the last five decades, percolation theory has brought new understanding to a broad range of topics in networks science \cite{boccaletti2006complex,castellano2009statistical,newman2010networks} and epidemiology \cite{moore2000epidemics,fu2008epidemic}. During 1959 to 1961, Erd{\"o}s and R{\'e}nyi proposed the evolution
model of random graphs, i.e., Erd{\"o}s-R{\'e}nyi (ER) model \cite{erdds1959random,erd6s1960evolution}. In this model, a randomly selected link is added to the network at each time step. Such random edge selection rule generates a second-order percolation transition that the largest connected component grows continuously. In 2009, Achlioptas introduced a product rule of edge selection \cite{achlioptas2009explosive}. In the model, two candidate edges are randomly selected, and the one
merging the two components with smaller product of their sizes is added to the network.
It postpones the emergence of the giant connected component and hence leads to a abrupt jump of the largest component size, which is called as explosive percolation. It is in striking contrast to the second-order percolation transition in the classical ER model, and a great deal of researches have been triggered
\cite{ziff2009explosive,radicchi2009explosive,cho2009percolation,friedman2009construction,
riordan2011explosive,lee2011continuity,araujo2010explosive,cho2013avoiding,chen2011explosive,d2015anomalous}. In
particular, refs. \cite{cho2009percolation,radicchi2009explosive} explored the explosive percolation of the configuration scale-free networks in a cooperative Achlioptas growth process. Ref. \cite{cho2009percolation}
found that there exists a critical degree-exponent parameter $\lambda_c\in(2.3 , 2.4)$. If $\lambda<\lambda_c$, the transition threshold tends to zero in the thermodynamic limit, and the transition is continuous; otherwise, a
discontinuous percolation transition occurs near a finite threshold. Almost at the same time, ref. \cite{radicchi2009explosive} got similar results through numeric simulations. However, it is proven rigorously that the explosive percolations for all the Achlioptas processes are actually continuous by subsequent researches
\cite{da2010explosive,grassberger2011explosive,lee2011continuity}.


Besides the Achlioptas process, another percolation model was introduced by Bohman, Frieze and Wormald (BFW)
\cite{bohman2004avoidance}. In this model, a single edge is randomly selected at each step. If the size of the
resulting component is less than a specified value, this edge would be added to the network, or a critical function is used to decide whether to receive it or not. Later on, Chen \emph{et al}. \cite{chen2011explosive} demonstrated that the transition of the BFW model is actually discontinuous.

Although percolation transition in random networks has been studied extensively, its application to real-world
networks is still scarcely investigated \cite{rozenfeld2010explosive,pan2011using}. Actually, in the evolution processes of most real-world networks, the strategy of selecting a newly added edge follows the mechanism of preferential selection instead of random selection. The preferential attachment was introduced by Barab{\'a}si and Albert \cite{barabasi1999emergence}, and it is considered as a vital mechanism to describe the evolution of the real-world networks with heterogeneous architecture \cite{newman2001clustering,de2007preferential,holme2002growing,small2015growing}. In the model,
the probability that a new node leads an edge to an exist node $i$ is in proportion to its degree $k_i$,
that is $\prod_{i}\sim k_i$. However, the architectures of most real-world networks can not be described by strictly
scale-free or random model \cite{newman2001random,fararo1964study}. Thus
a more realistic network evolution model with hybrid attachment was proposed \cite{liu2002connectivity,liu2003propagation}.
In the model, the probability that a node $i$ is attached by a new edge is in proportion to $\prod_{i}\sim(1-q)k_i+q$, where
$q$ controls the relative weights between the heterogeneous and homogeneous components in the evolution process.


In this paper, we propose a hybrid BFW model (HBFW) with mixture of preferential and random edge selection to study the network percolation in a more realistic case. When $q=0$, the model degenerates to the original BFW model, where each candidate edge is selected randomly, and the degrees thus satisfy exponential distribution. When $q=1$, the network is
strictly scale free as the preferential attachment dominates network evolution process. Simulation results demonstrate that three
separate parameter regions $\left[0,q^{\uppercase\expandafter{\romannumeral1}}_c
\right]$, $\left(q^{\uppercase\expandafter{\romannumeral1}}_c,q^{\uppercase\expandafter{\romannumeral2}}_c \right]$,
$\left(q^{\uppercase\expandafter{\romannumeral2}}_c, 1\right]$ exist in the model,
where $q^{\uppercase\expandafter{\romannumeral1}}_{c}\in (0.84, 0.86)$ and
$q^{\uppercase\expandafter{\romannumeral2}}_c\in (0.97, 0.98)$ respectively stand for two critical values. And crossover
phenomena of percolation transition occur between these regions. Specifically, there is only a single discontinuous transition when $q\in\left[0, q^{\uppercase\expandafter{\romannumeral1}}_c \right]$. And when
$q\in\left(q^{\uppercase\expandafter{\romannumeral1}}_c,q^{\uppercase\expandafter{\romannumeral2}}_c \right]$, multiple
discontinuous jumps of the largest component size occur in the evolution process. Moreover,
when $q\in\left(q^{\uppercase\expandafter{\romannumeral2}}_c, 1\right]$, the first transition of the HBFW model becomes continuous. Most extremely, there is single giant component all through the process, when $q$ approaches to 1. It shows totally different properties of percolation transitions in a network with more realistic evolution mechanism,
compared with those in random networks or configuration networks.

\section{PERCOLATION MODEL} \label{sec:model}
In our model, we use the mechanism of the hybrid attachment to modify the rule of selecting a candidate edge in the BFW model. By introducing a tunable parameter \emph{q}, we get a hybrid rule of preferential selection and random selection.

Initially, the system is a network \emph{G} with \emph{N} isolated nodes. Denote \emph{K} as the cap size and set $\emph{K=2}$. $u\left(l\right)$ and $L\left(l\right)$ respectively represent the number of candidate edges and accepted edges at the current step \emph{l}. And $t=L(l)/N$ is the density of accepted edges at the current step. Then we consider the evolution process of the HBFW model. Namely, at each step \emph{l}, the two nodes attaching to a candidate edge $e_u$ is selected according to the following two steps:

$1)$ The first node is selected randomly from the system.

$2)$ The second node is selected by the rule of hybrid attachment. The node $i$ is selected by the preferential (random) selection strategy with probability $q$ ($1-q$).

According to the selection rule described above, a candidate edge $e_u$ is obtained, and then it is judged by the BFW algorithm.

$3)$ Edge addition in the BFW algorithm. We can decide whether the candidate $e_u$ should be accepted or not according to the following criteria:

(\textbf{\emph{i}}) Denote $C_1$ as the size of the largest component if $e_u$ is added to the system;

(\textbf{\emph{ii}}) If $C_1\leq K$, we accept $e_u$ , and set $L(l+1)=L(l)+1$, $u(l+1)=u(l)+1$, $l=l+1$;

(\textbf{\emph{iii}}) Otherwise, if $L(l)/u(l)<g(K)=1/2+(2K)^{-1/2}$, set $K=K+1$, and go to (\emph{ii}), until $C_1\leq K$ or $L(l)/u(l)>g(K)$. If $C_1\leq K$, $e_u$ is added to the network, and set $L(l+1)=L(l)+1$, $u(l+1)=u(l)+1$, $l=l+1$; or if $L(l)/u(l)>g(K)$, $e_u$ is rejected, and set $u(l+1)=u(l)+1$, $l=l+1$;

Repeat steps $1) \sim 3)$. Note that initially all the nodes in the system are isolate, so the selection rule of
preferential attachment can not be applied to any node. Thus without loss of generality, at the first step, the two nodes are randomly selected.

\section{SIMULATION RESULTS} \label{sec:simulation}
\begin{figure}
\begin{center}
\includegraphics[height=80mm,width=58mm]{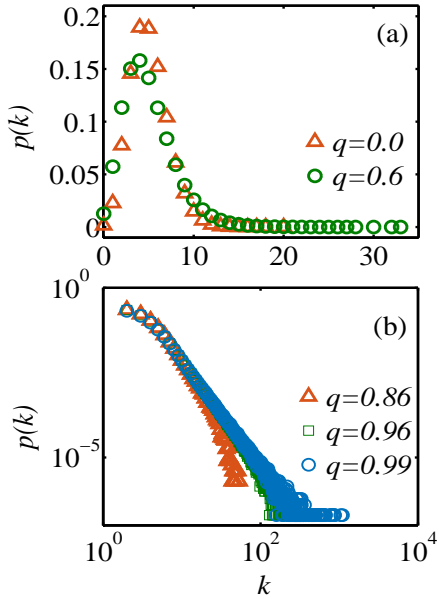}
\caption{(Color online) Typical degree distribution of generated networks with $\langle k\rangle =5$. (a) the degree distribution for $q=0.0$ and $q=0.6$, (b) the degree distribution for $q=0.86$, $q=0.96$, $q=0.99$. System size is set sa $N=10^6$.}
\label{digDist(mix)}
\end{center}
\end{figure}

First of all, we explore the effect of the hybrid attachment on the architecture of the generated networks. In order to demonstrate the result clearly and without loss of generality, we set the generated networks of
$\langle k \rangle =5$. We exhibit the degree distribution of the networks with five typical values of parameter $q$,
as shown in Figs.~\ref{digDist(mix)} (a) and (b).  When \emph{q} is relatively small (e.g., $q=0.0$ and $q=0.6$), a pair
of nodes connected by a candidate edge are selected almost by the strategy of random selection, thus degrees of all nodes
satisfy an exponential distribution \cite{newman2001random} [see fig. ~\ref{digDist(mix)} (a)]. With the increase of $q$, it is more likely to select a
large-degree node, then the heterogeneity of the networks grows gradually. And after about $q=0.86$, the degree
distribution of the networks evolves asymptotically to power-law distribution \cite{de2007preferential}, as shown in
Fig.~\ref{digDist(mix)}(b).

Next, we discuss the properties of percolation transition in the HBFW model. For clarity, we denote
$NC_i \thicksim o(N)$ if $\lim_{N\rightarrow\infty}C_{i}\rightarrow 0$, where $C_i$ is the fraction of nodes in the
\emph{ith} largest component, and $C_1$ is defined as the order parameter of network percolation. If
$\lim_{N\rightarrow\infty}C_{i}\rightarrow c, c\in \left(0,1 \right]$, we denote $NC_i\thicksim O \left( N \right)$.

\begin{figure}
\begin{center}
\includegraphics[width=1\linewidth]{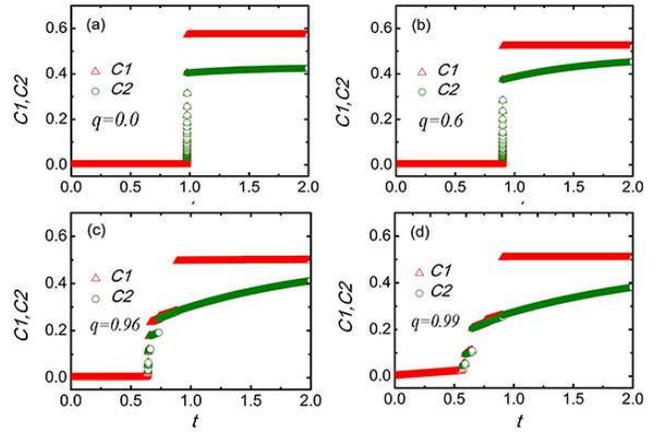}
\caption{(Color online) Time evolutions of $C_1$ and $C_2$ for four typical values of $q$. (a) and (b) respectively depict the evolution processes for $q=0.0$ and $q=0.6$, which are both smaller than $q^{\uppercase\expandafter{\romannumeral1}}_c$; (c) and (d) respectively depict the processes for $q=0.96$ and $q=0.99$ , which are both larger than $q^{\uppercase\expandafter{\romannumeral1}}_c$. System size is set as $N=10^6$.
}
\label{orderparam}
\end{center}
\end{figure}

In order to investigate the impact of hybrid attachment on percolation transition, we perform simulations with varieties of $q$. In the evolution process, edges are selected in strict accordance with the rules mentioned above. We first focus on the time evolutions of $C_1$ and $C_2$. Fig.~\ref{orderparam} illustrates the effect of the hybrid attachment on the percolation transition of the HBFW model by exhibiting the results for $q=0.0, q=0.6, q=0.96$ and $q=0.99$. Fig.~\ref{orderparam} shows that when $q=0.0$, it degenerates to the BFW model and there is only a unique discontinuous jump of the largest component at the percolation threshold. But for $q>0$, as shown in Figs.~\ref{orderparam}(b)$\sim$(d), we can see that the transition point decreases gradually with the probability $q$ of preferential attachment. Most importantly,
for a large value of $q$ multiple jumps of $C_1$ and $C_2$ occur after the first percolation transition [see Figs.~\ref{orderparam}(c) and (d)]. When $q\rightarrow1$, as shown in Fig.~\ref{orderparam}(d), the first transition of $C_1$ and $C_2$ is prone to be continuous.

From large numbers of simulations, we find that there exists two critical values
$q^{\uppercase\expandafter{\romannumeral1}}_c\in (0.84, 0.86)$ and
$q^{\uppercase\expandafter{\romannumeral2}}_c\in (0.97, 0.0.98)$. For
$q\in[0,q^{\uppercase\expandafter{\romannumeral1}}_c]$, there is only one jump of $C_1$ and $C_2$ at the transition point.
And for $q\in(q^{\uppercase\expandafter{\romannumeral1}}_c, q^{\uppercase\expandafter{\romannumeral2}}_c]$, multiple jumps
of the $C_1$ and $C_2$ emerge. However, the first transition
of $C_1$ and $C_2$ is prone to be continuous for $q\in(q^{\uppercase\expandafter{\romannumeral2}}_c, 1]$.
In the following section, we focus on the time evolution of the largest component, and demonstrate the above conclusions from
two aspects. Firstly, by analyzing the jump size of the largest component $\Delta C_1$ \cite{nagler2011impact}, and the time
interval $\Delta t_i, i=1...3$ between two adjacent jumps of the largest component, we demonstrate the existence of
multiple discontinuous jumps for
$q\in(q^{\uppercase\expandafter{\romannumeral1}}_c, q^{\uppercase\expandafter{\romannumeral2}}_c]$. And we strengthen the above results by investigating the susceptibility $\chi$ of $C_1$. Secondly, by using the technology of ``critical window''
\cite{achlioptas2009explosive}, we examine whether the first transition is discontinuous or not for different value of $q$.

\begin{figure}
\begin{center}
\includegraphics[width=1\linewidth]{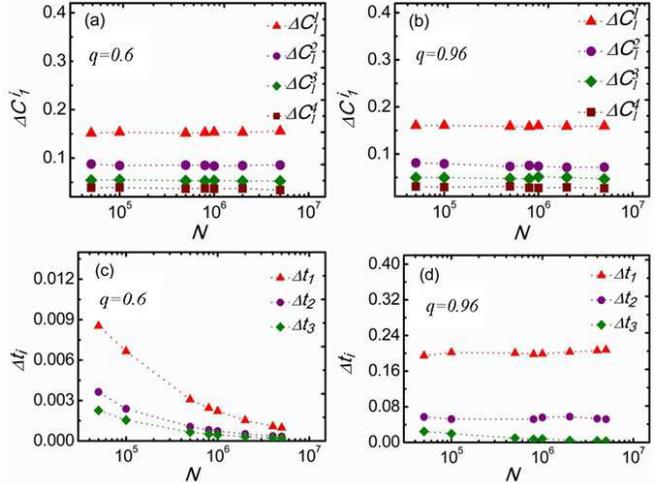}
\caption{(Color online) Finite size effects of multiple jumps of the largest component
 $\Delta{C_1}$ and the time intervals between each two adjacent jumps $\Delta{t_1}\sim\Delta{t_3}$. The four largest jumps of order parameter $\Delta{C^{1}_{1}}\sim \Delta{C^{4}_{1}}$ for $q=0.6$ (a) and $q=0.96$ (b). Time delays $\Delta{t_1}\sim\Delta{t_3}$ for $q=0.6$ (c) and $q=0.96$ (d). Data are averaged over 100 network realizations.}
\label{jump}
\end{center}
\end{figure}

Firstly, we demonstrate that there is only one discontinuous jump of
$C_1$ for $q\in[0, q^{\uppercase\expandafter{\romannumeral1}}_c]$, and multiple discontinuous jumps for
$q\in(q^{\uppercase\expandafter{\romannumeral1}}_c, q^{\uppercase\expandafter{\romannumeral2}}_c]$. In fact, Ref.
\cite{nagler2011impact} pointed out that if the maximum increase of  the order parameter $\Delta C^{max}_{1}$ caused by
adding a single edge satisfies the condition
\begin{equation}\label{maxJump}
  \lim_{N\rightarrow \infty}\Delta C^{max}_{1}/N>0,
\end{equation}
the transition is strongly discontinuous. We consider the finite size effects of top 4 largest jumps of the largest component for different
values of $q$. Figs.~\ref{jump} (a) and (b) show the cases of $q=0.6$ and $q=0.96$, where $\Delta C^{i}_1$ stands for the
$ith$ largest jump size of the largest component. Simulations results illustrate that the size of the jumps are
independent of the system size $N$. Therefore, the top four largest jumps of the largest component is discontinuous
for $q\in[0,q^{\uppercase\expandafter{\romannumeral1}}_c]$ in the thermodynamic limit. Next, we
investigate the time delays between two adjacent jumps of the largest component to judge how many discontinuous jumps occur in
the evolution process. The $\Delta t_i, i=1,2,3$ stands for the time delay between the $ith$ jump and the$(i+1)th$ jump.
Fig.~\ref{jump}(c) shows that when $q=0.6$, $\Delta t_i(i=1,2,3)$ converge to zero in the thermodynamic limit. This means
that these discontinuous jumps in the HBFW process with $q=0.6$ occur at almost the same time when \emph{N} is large
enough. Thus there is only one discontinuous jump of $C_1$ when $N\rightarrow\infty$. For $q=0.96$, the
two time delays converge asymptotically to two positive constant values in the thermodynamic limit, i.e. $\lim_{N\rightarrow \infty}\Delta t_{1}=0.206\pm 0.002$ and $\lim_{N\rightarrow \infty}\Delta t_{2}=0.057\pm 0.004$  [see Fig.~\ref{jump}(d)]. This means that three separate discontinuous jumps of $C_1$ occur during the percolation process.

\begin{figure}
\begin{center}
\includegraphics[width=1\linewidth]{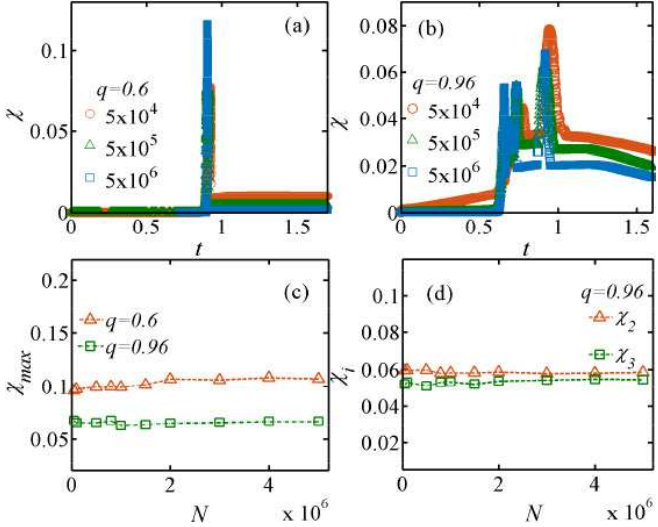}
\caption{(Color online) Time evolution and finite size effects of the susceptibility $\chi$. Time evolution of the $\chi$ for $q=0.6$ (a) and $q=0.96$ (b). (c) The maximum value of susceptibility $\chi_{max}$ versus network size for $q=0.6$ and $q=0.96$. (d) The other two peak values of susceptibility $\chi_{2}$ and $\chi_{3}$ versus network size for $q=0.6$ and $q=0.96$. Data are averaged over 100 network realizations.}
\label{deviation}
\end{center}
\end{figure}

To strengthen the above conclusions, we consider the susceptibility
\begin{equation}\label{chi}
  \chi=\sqrt{\langle C^2_1\rangle-\langle C_1\rangle^2},
\end{equation}
which quantifies the amplitude of the fluctuations of the
largest component size \cite{shu2015numerical}. When a transition occur in the evolution process, $\chi$ reaches to a peak value $\chi_{max}$.
Further, a none zero value of $\chi_{max}$ in the thermodynamic limit indicates a discontinuous transition
\cite{binder1981finite,voss1984fractal}. In Fig.~\ref{deviation}, we exhibit the susceptibility of the largest component
sizes for $q=0.6$ and $q=0.96$ with corresponding to Fig.~\ref{jump}. Fig.~\ref{deviation}(a) shows that there is only one
peak of $\chi$, and Fig.~\ref{deviation}(b) tells us that it has three obvious peaks of $\chi$ as time \emph{t} increases.
For the sake of clarity, we denote the second and third largest value of $\chi$ for $q=0.96$ as $\chi_{2}$ and $\chi_{3}$.
Results of Figs.~\ref{deviation}(a) and (b) show that for $q=0.96$, there are three jumps of the largest component size, but for
$q=0.6$ only one jump of the largest component size occurs. Besides, Figs.~\ref{deviation}(c) and (d) demonstrate that the
three peak values of the susceptibility for $q=0.96$, and the maximum value of the susceptibility for $q=0.6$ tend to be
nonzero values in the thermodynamic limit. Thus, it leads to the conclusion that the jump of the largest component for
$q=0.6$ and the three jumps for $q=0.96$ are essentially discontinuous.
\begin{figure}
\begin{center}
 \includegraphics[height=86mm,width=55mm]{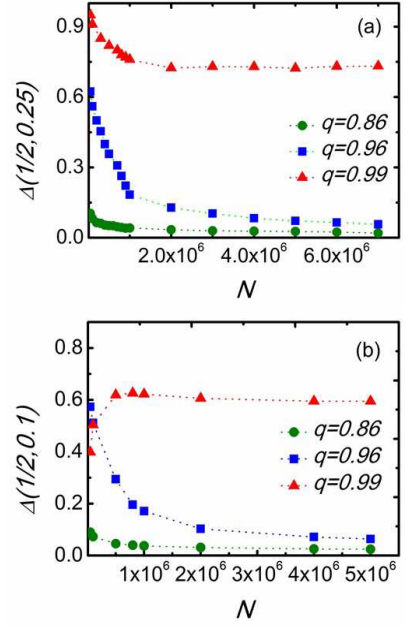}
\caption{(Color online) Relationship between the rescaled size of critical window $\Delta(\alpha,\beta)$ and network size \emph{N}. (a) $\alpha=1/2,\beta=0.25$ and (b) $\alpha=1/2,\beta=0.1$ are chosen in $\Delta(\alpha,\beta)$ for $q=0.86,q=0.96$, and $q=0.99$. Data are averaged over 100 realizations.}
\label{cw}
\end{center}
\end{figure}

At last, we demonstrate the continuity of the first transition for $q\in(q^{\uppercase\expandafter{\romannumeral2}}_c, 1]$
in the HBFW model by applying the theory of ``critical window'' proposed in Ref. \cite{achlioptas2009explosive}. Let $L_0$
denote the number of added edges at the last moment for
\begin{equation}\label{L0}
  NC_1\leq N^\alpha,
\end{equation}
 and $L_1$ as the number of added edges at the first moment for
  \begin{equation}\label{L1}
    NC_1\geq \beta N
  \end{equation}
where $\alpha,\beta \in(0,1)$. $\Delta (\alpha,\beta)=L_1-L_0$ denotes the number of added edges during the evolution process, which is called as critical window. Fig.~\ref{cw} shows the rescaled value of $\Delta(\alpha, \beta)/N$ as a function of system size \emph{N} for $q=0.86$, $q=0.96$, and $q=0.99$. We first set $\alpha=1/2, \beta=0.25$. As shown in
Fig.~\ref{cw}(a), $\Delta(\alpha, \beta)/N$ is sublinear to $N$ and satisfies
$\lim_{N\rightarrow \infty}\Delta(\alpha, \beta)/N=0$ for $q=0.86$ and $q=0.96$, while
$\lim_{N\rightarrow \infty}\Delta(\alpha, \beta)/N=0.732\pm0.002$ for $q=0.99$ [see Fig.~\ref{cw}(a)]. For more accurate,
we lower the value of $\beta$. In Fig.~\ref{cw}(b), we set $\beta=0.1$ and get the accordant result as
Fig.~\ref{cw}(a). Figs.~\ref{cw}(a) and (b) reveal the fact that the first jump of the HBFW process is
discontinuous for $q=0.86$ and $q=0.96$, but it is continuous for $q=0.99$. More simulations demonstrate the existence of
the critical value $q^{\uppercase\expandafter{\romannumeral2}}_c\in(0.97,0.98)$, below which the first jump of the largest component in HBFW process is discontinuous, but above which it is continuous.
\begin{figure}
\begin{center}
\includegraphics[height=120mm,width=65mm]{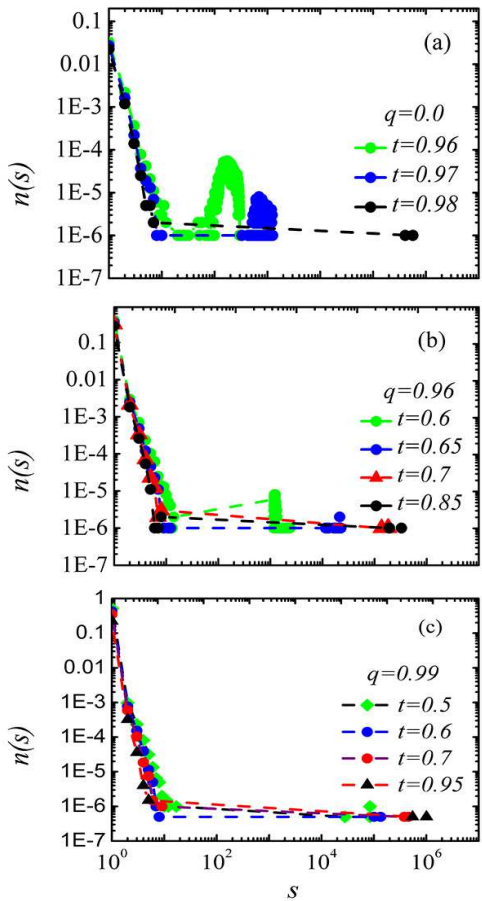}
\caption{(Color online) The component distributions at the vicinity of transition threshold for $q=0.0$ (a), $q=0.96$ (b) and $q=0.99$ (c). System size is $N=10^6$.}
\label{cluster}
\end{center}
\end{figure}

Finally, we give a qualitative analysis for the two crossover phenomena of percolation transitions in the HBFW model. In
the BFW model, the evolution of the largest component is restricted by the cap size $K$. Thus at each time step, the small
components are merged together by a newly added edge with high probability. Consequently, a mass of components, whose
sizes are approximately equal to the largest component size, emerge in the subcritical regime. In other words, a ``powder key'' is formed,
which is a necessary condition for the discontinuous percolation transition \cite{friedman2009construction}. As shown in
Fig.~\ref{cluster} (a), there are a mass of components with size of about two hundred nodes at $t=0.96$, and with size of
about seven hundred nodes at $t=0.97$. Subsequently, a vanishingly small number of added edges can merge these small
components, and thus lead to an explosive percolation. And at $t=0.98$, two giant components
formed.

The hybrid attachment mechanism in the HBFW model makes evolution procedure much more complex than the classical BFW
process, which leads to peculiar properties of the percolation transitions. Generally, there are two underlying competed
mechanisms behind the percolation procedures in the HBFW model, which are the suppression and promotion of growth of the
largest component, respectively. On the one hand, the cap size $K$ restricts the formation of large components. On the other
hand, with the increase of $q$, the probability of selecting large-degree nodes rises, which accelerates the
growth of large components. When $q<q^{\uppercase\expandafter{\romannumeral1}}_c$, the strategy of selecting a candidate
node is predominated by the random selection [see Fig.~\ref{digDist(mix)} (a)], the suppression plays a main role,
rendering the percolation type similar to BFW model [see Figs.~\ref{orderparam} (a) and (b)]. With the increase of $q$,
edges are more likely to attach to large-degree nodes [see Fig.~\ref{digDist(mix)} (b)]. Owing to the suppression of
cap size \emph{K}, the large-degree nodes absorb more and more small components or isolate nodes into the components, to
which these hub nodes belong.
Consequently, a mass of hub-centric components with sizes being sublinear to the system size $N$ appear in the subcritical regime.
As shown in Fig.~\ref{cluster} (b), there are many components with size of about one thousand nodes at $t=0.6$. In the subsequent procedure,
each added edge would combine two of these hub-centric components with a very small probability, and thus leads to a discontinuous jump of the largest
component. As shown in Fig.~\ref{cluster}, three discontinuous jumps of the largest component size occur in the time duration
$[0.6,0.65]$, $(0.65,0.7]$ and $(0.7,0.85]$.


When $q$ is large enough, i.e., $q>q^{\uppercase\expandafter{\romannumeral2}}_c$, the cap value $K$ that constrains the
growth of the largest component won't work anymore. Affected by the mechanism of preferential attachment, the large-degree nodes absorb small components or isolate nodes continuously, and giant components form at the early stage of the evolution process. Thus, it leads to the continuous growth of the largest component. As shown in Fig.~\ref{cluster} (c), several components with size of about ten thousand nodes form at $t=0.5$.
We see that the large components absorb more small components, thus several giant components form at $t=0.6$. In the following time steps, these
giant components are merged by edges, and at last the two stable giant component exist in the supercritical regime, e.g., $t=0.7$ and $t=0.95$.
If $q$ approaches to 1, at each time step, the probability of selecting the second node approximates to $\prod_{i}\sim k_i$. The
isolate nodes or small components will be absorbed into the large components continuously by the large degree nodes. And when $q=1$, we can
imagine that there will be a single giant component to which those large-degree nodes belong, and it grows continuously all through the process.
Thus, there will be no jump of the largest component.


\vspace{10pt}

\section{Discussion} \label{sec:dis}
In this paper, we propose a network percolation model with the hybrid attachment based on the BFW process. In the model,
the tune parameter $q$ controls the evolution strategy. When $q$ is relatively small, the strategy of random edge selection
dominates the evolution process. When $q$ becomes larger, the strategy of preferential edge selection
dominates the evolution process. A large number of numeric simulations indicate that there exist crossover phenomena of
percolation transition between three separate regions
$\left[0,q^{\uppercase\expandafter{\romannumeral1}}_c \right]$, $\left(q^{\uppercase\expandafter{\romannumeral1}}_c,q^{\uppercase\expandafter{\romannumeral2}}_c \right]$,
$\left(q^{\uppercase\expandafter{\romannumeral2}}_c, 1\right]$. In the region $q\in[0,q^{\uppercase\expandafter{\romannumeral1}}_c]$, the strategy of random edge selection
dominates the evolution process, and with the constraint of forming the giant component, there is only one discontinuous
jump of the largest component at the percolation threshold as the origin BFW model. In the region
$q\in(q^{\uppercase\expandafter{\romannumeral1}}_c,q^{\uppercase\expandafter{\romannumeral2}}_c]$, the strategy of more
preferential edge selection dominates the evolution process, and multiple discontinuous jumps occur as a result of competition
between the mechanisms of constraining and accelerating the forming of the giant component, which are caused by the
restricted of the cap size $K$, and the preferential selection of a candidate edge respectively. Further, in the region
$q\in(q^{\uppercase\expandafter{\romannumeral2}}_c,1]$, with nearly complete preferential selection, the mechanism of
constraining the giant connected component is no longer function, the largest component grows continuously at the first
transition point. And if $q$ approaches to 1, a single giant component grows continuously throughout the evolution
process.

Percolation of networks has been one of the hottest topics in recent years, and it is considered as an important tool
to study the robustness of networks \cite{parshani2010interdependent}, spreading of epidemics \cite{newman2002spread} and so
on. Although many important and interesting percolation phenomena are observed on both random networks and lattices,
it is the first time to study the percolation in network evolution process with hybrid attachment, which will help us to
understand the percolation transition in a more realistic process. It is unusual to find the crossover phenomena of
percolation transition in the network evolution process, which have recently stimulated broad attention in other fields,
such as the processes of social contagions \cite{wang2016dynamics}. Specially, a recent work studied the percolation
transition in scale-free networks and the multiple discontinuous percolation transitions are observed
\cite{chen2015multiple}. We study the percolation of networks in a more realistic evolution process from the
aspect of numeric simulations, but its theoretical analysis need to be carried out. Besides, the percolation transition in
real-world networks needs further works.

\acknowledgments

This work was partially supported by the National Natural Science Foundation of China under Grants Nos. 11105025, 11575041 and 61433014, and Project No. 9140A06030614DZ02083.


\end{document}